# AMV-L: Lifecycle-Managed Agent Memory for Tail-Latency Control in Long-Running LLM Systems

Emmanuel Bamidele
Department of Computer Science, Georgia Institute of Technology

**Abstract**

Long-running LLM agents require persistent memory to preserve state across interactions, yet most deployed systems manage memory with age-based retention (e.g., TTL). While TTL bounds item lifetime, it does not bound the computational footprint of memory on the request path: as retained items accumulate, retrieval candidate sets and vector similarity scans can grow unpredictably, yielding heavy-tailed latency and unstable throughput. We present AMV-L (Adaptive Memory Value Lifecycle), a memory-management framework that treats agent memory as a managed systems resource. AMV-L assigns each memory item a continuously updated utility score and uses value-driven promotion, demotion, and eviction to maintain lifecycle tiers; retrieval is restricted to a bounded, tier-aware candidate set that decouples the request-path working set from total retained memory. We implement AMV-L in a full-stack LLM serving system and evaluate it under identical long-running workloads against two baselines: TTL and an LRU working-set policy, with fixed prompt-injection caps. Relative to TTL, AMV-L improves throughput by 3.1× and reduces latency by 4.2× (median), 4.7× (p95), and 4.4× (p99), while reducing the fraction of requests exceeding 2s from 13.8% to 0.007%. Compared to LRU, AMV-L trades a small regression in median/p95 latency (+26% / +3%) for improved extreme-tail behavior (−15% p99; −98% >2s) and lower token overhead (≈6% fewer tokens/request), while matching retrieval quality (value means within ≈0–2%). The gains arise primarily from bounding retrieval-set size and vector-search work, not from shortening prompts. Our results show that predictable performance for long-running LLM agents requires explicit control of memory working-set size and value-driven lifecycle management, rather than retention time alone.

## 1. Introduction

LLM-based agents are increasingly deployed as long-running services that must preserve state across interactions. Personal assistants accumulate user preferences, coding agents retain project context, and autonomous services maintain intermediate task state over extended executions. In these systems, persistent memory is necessary for coherence, continuity, and efficiency, but it also becomes a dominant performance variable once memory participates in the request path[1,2].

Most deployed agent stacks manage memory with age-based retention, typically time-to-live (TTL). TTL is operationally convenient: it bounds how long an item remains stored and yields predictable storage growth under steady ingest[3]. However, TTL does not bound the *computational footprint* of memory[4]. In LLM agents, retained items are not inert: they influence retrieval candidate-set size, vector similarity-search cost, prompt construction overhead, and ultimately end-to-end inference latency. Consequently, TTL can keep storage age bounded while allowing request-path compute to grow without a corresponding bound.

The core systems issue is the mismatch between *total retained memory* and the *effective working set* exercised during request processing. Under TTL, large numbers of items may remain simultaneously eligible for retrieval. Even when median behavior appears healthy, rare requests can activate unusually large candidate sets and vector scans, producing heavy-tailed latency. These outliers dominate capacity planning and user experience because production services are constrained by p95/p99 latency and tail SLOs, not by mean or median latency.

We argue that agent memory must be treated as a managed systems resource rather than a passive store. Analogous to how operating systems regulate CPU and physical memory via explicit policies and budgets, LLM agents require memory policies that regulate not only *retention*, but also the per-request *compute* induced by memory. A natural baseline for bounding compute is an LRU working-set policy, which reduces retrieval footprint by prioritizing recency. However, LRU is value-agnostic: it can evict long-lived but high-utility information under phase shifts, and it provides limited control over the interaction between utility, eligibility, and retrieval cost, precisely the axes that determine tail behavior in long-running workloads.

AMV-L (Adaptive Memory Value Lifecycle) addresses this problem by coupling value-aware retention with explicit working-set control. AMV-L assigns each memory item a continuously updated utility score and manages memory through value-driven promotion, demotion, and eviction across lifecycle tiers. Retrieval is restricted to a bounded, tier-aware candidate set, decoupling the request-path working set from total retained memory while preserving high-utility long-term knowledge.

We implement AMV-L in a production-style, full-stack LLM serving system and evaluate it under identical long-running workloads against two baselines: TTL and LRU, with fixed prompt-injection caps. Relative to TTL,

AMV-L improves throughput by 3.1× and reduces latency by 4.2× (median), 4.7× (p95), and 4.4× (p99), while reducing the fraction of requests exceeding 2s from 13.8% to 0.007%. Compared to LRU, AMV-L trades a small regression in median/p95 latency (+26% / +3%) for improved extreme-tail behavior (−15% p99; −98% >2s) and lower token overhead (≈6% fewer tokens/request), while matching retrieval quality (value means within ≈0–2%). The improvements arise primarily from bounding retrieval-set size and vector-search work, not from prompt compression.

Together, these results identify uncontrolled memory working-set growth as a primary driver of tail latency in long-running LLM agents, and show that lifecycle-aware, value-driven memory management can restore predictable performance beyond what age-based retention or recency-only policies provide.

In summary, this paper:

a) Formulates persistent agent memory as a *request-path systems resource* whose working set must be explicitly bounded to control tail latency.

b) Introduces AMV-L, a value-driven lifecycle policy that decouples request-path eligibility from total retained memory via tiered management and bounded retrieval.

c) Demonstrates large tail-latency and throughput gains over TTL, and a clear tradeoff frontier versus LRU: slightly worse median/p95, but better extreme-tail and lower token cost at comparable retrieval quality.

## 2. Background and Motivation

Modern LLM-based agents maintain persistent memory to support long-running behavior [2,5–8]. Memory records include user facts and preferences, summaries of prior interactions, intermediate task artifacts, retrieved documents, and system-generated observations [9–13]. During request processing, an agent retrieves a subset of these items and incorporates them into the context presented to the model, either directly or through a retrieval-augmented generation pipeline [14–17]. This design couples memory to the request critical path [18–20]. The set of memory items considered for retrieval determines embedding generation, similarity-search work, candidate selection and reranking cost, and the amount of context assembled for inference [21–23]. As a result, memory management is not only a storage problem. It is a primary determinant of latency, throughput, and serving cost [24–26].

Most agent frameworks manage memory using time-based retention, typically time-to-live (TTL) [3,27,28]. Under TTL, each item is assigned an expiration time and is deleted once it becomes sufficiently old [29]. TTL is attractive because it is simple to implement and reason about operationally. It bounds item age and can bound storage growth under stationary write rates [30]. However, TTL is indifferent to utility and indifferent to computational footprint. Items that differ substantially in usefulness or retrieval cost are treated equivalently as long as they fall within the retention window [31,32]. More importantly, TTL does not directly control which stored items remain eligible to participate in request processing.

This motivates a distinction between total retained memory and the retrieval working set. Total retained memory is the number of items stored by the system. The retrieval working set is the subset that is eligible for similarity search and candidate selection on a given request. Only the working set directly contributes to request-path computation [33–35], yet many deployments implicitly equate retention with eligibility [36,37]. Under TTL, the eligible pool can grow with the retained store because all non-expired items remain candidates. Even when prompt injection is bounded by a fixed top-$n$ cap, the system may still have to search across a large eligible pool in order to identify those $n$ items. Prompt caps bound the final prompt length. They do not bound similarity-search work [38,39].

This mismatch between retention and eligibility creates heavy-tailed behavior. For many requests, the system searches a modest eligible pool and returns quickly. For a minority of requests, the eligible set is much larger or the search work is higher, producing latency outliers that dominate operational behavior [40]. Service-level objectives are typically expressed in high percentiles, so rare slow requests can determine provisioned capacity, queueing behavior, and perceived reliability [41,42]. In our evaluation, TTL exhibits heavy-tailed latency and low throughput, while policies that reduce eligibility, including LRU and AMV-L, collapse the tail and increase throughput under the same injection cap. This confirms that eligibility control is the key mechanism for predictable performance in long-running agent memory.

The systems implication is that agent memory must be managed as a computational resource, not only a persistence layer. A practical policy should regulate the retrieval working set that participates in similarity search, preserve high-utility information over long horizons, and adapt to non-stationary access patterns [43,44]. Recency-based working-set policies such as LRU provide an effective baseline for bounding retrieval cost, but they are value-agnostic and can fail to retain long-lived high-utility items under phase shifts [45–47]. This motivates lifecycle-aware memory management that explicitly models utility and uses that signal to control eligibility. AMV-L is designed around this perspective. It assigns each memory item a continuously updated utility value and uses value-driven lifecycle transitions to control which items remain on the high-cost retrieval path, while retaining long-lived knowledge outside the default working set.

## 3. Design Goals

AMV-L is designed to replace age-based retention with a lifecycle policy that explicitly controls the computational footprint of memory. The design is guided by four goals.

<u>G1. Bound request-path memory cost independently of total retained memory.</u>

The primary goal is to bound the computational cost that memory imposes during request processing. In long-running agents, total stored memory may grow over time, but request latency and throughput must remain stable. AMV-L therefore regulates which items are eligible for retrieval and prompt construction, rather than only how many items are stored. The key requirement is that retrieval candidate-set size and similarity-search work remain bounded as total retained memory increases.

<u>G2. Preserve high-utility memory over long-time horizons.</u>

A policy that aggressively deletes old items can improve latency but can destroy long-term agent utility. AMV-L must preserve information that continues to provide value, such as user preferences, persistent task context, and domain facts, even when it is old. The system should support indefinite retention of high-utility items while suppressing low-utility items from the high-cost retrieval path.

<u>G3. Adapt to non-stationary access patterns.</u>

Memory utility changes over time. Frequently used items can become irrelevant, and previously unused items can later become important. AMV-L must continuously update per-item utility using online signals, such as access frequency, contribution to prompt construction, and recency, and adjust lifecycle state accordingly. The working set should track current utility without requiring offline retraining or global recomputation.

<u>G4. Impose low management overhead and integrate with existing agent stacks.</u>

Memory management must not introduce overhead that offsets the latency gains it is intended to provide. Value updates, tier transitions, and eviction decisions should be lightweight, incremental, and local to each item. AMV-L should integrate into existing retrieval-augmented agent pipelines with minimal interface changes, exposing standard operations for insertion, retrieval eligibility, and lifecycle maintenance.

Together, these goals define AMV-L as a working-set management mechanism for agent memory. It preserves long-term utility while enforcing predictable request-path computation.

## 4. AMV-L Overview

AMV-L (Adaptive Memory Value Lifecycle) manages agent memory as a working-set control mechanism. The system assigns each memory item a continuously updated scalar value that estimates utility and uses that value to determine two outcomes: the item's lifecycle state and whether it is eligible to participate in request-path retrieval. The design decouples total retained memory from the memory working set exercised by each request.

At a high level, AMV-L consists of three components: a value model, a tiered lifecycle, and a bounded retrieval path. The value model updates each item's utility score online using local usage signals. The lifecycle maps value to a small number of tiers (hot, warm, cold) and performs promotion, demotion, and eviction as values change. The retrieval path consults only a bounded subset of tiers, ensuring that similarity-search and prompt-construction work remain controlled even as total memory grows.

### 4.1 Memory items and value state

AMV-L stores memory as discrete items, for example facts, summaries, observations, or task artifacts. Each item m is represented by content, metadata, an embedding (if applicable), and a scalar value $V(m)$ that summarizes its current utility to the agent. Utility is not treated as static. AMV-L updates $V(m)$ incrementally as the system observes accesses, successful use in prompt construction, and inactivity over time.

A key constraint is that value updates are local and lightweight. AMV-L does not require global re-ranking of all memory or periodic full-database scans. Instead, each item carries sufficient state to update its value online, keeping management overhead small relative to inference and retrieval cost.

### 4.2. Tiered lifecycle organization

AMV-L partitions memory into a small set of lifecycle tiers:

a) Hot tier: items eligible for normal request-path retrieval and prompt construction.

b) Warm tier: retained items with moderate utility, excluded from the default high-frequency retrieval path.

c) Cold tier: low-utility items retained at minimal computational cost and excluded from normal retrieval.

These tiering separates retention from eligibility for computation. An item may remain stored in warm or cold without imposing cost on every request. In the default configuration, only hot-tier items contribute to the retrieval candidate set used during request processing. Warm and cold act as lower-cost reservoirs from which items may later be promoted if access patterns indicate renewed utility.

### 4.3 Lifecycle transitions

AMV-L moves items between tiers as their value changes. When an item is accessed or contributes to a request, its value increases and it may be promoted, for example warm to hot. When an item is not reused, decay lowers its value and it may be demoted, for example hot to warm or warm to cold. Items that remain low value in cold may be evicted.

Transitions are applied incrementally and asynchronously. Request processing records usage events and updates item value, but tier migration and cleanup are performed off the critical path. This prevents lifecycle maintenance from increasing request latency and allows AMV-L to adapt the working set continuously without blocking foreground traffic.

### 4.4 Bounded retrieval and prompt construction

The primary systems effect of AMV-L is on retrieval. Under TTL-style retention, many items remain simultaneously eligible for retrieval, so candidate-set size can grow with the number of retained items. A recency-based working-set policy such as LRU can reduce this footprint by preferentially keeping recently accessed items eligible, but it remains value-agnostic and can discard long-lived, high-utility information during phase shifts or long gaps. AMV-L instead bounds retrieval eligibility while explicitly modeling utility, allowing high-value items to persist outside the default retrieval path and re-enter it when their utility increases.

In AMV-L, retrieval is restricted to a bounded, tier-aware candidate set derived from the hot tier (and optionally a small warm-tier budget for recall-sensitive workloads). The retrieved set is then truncated by a fixed prompt-injection cap before prompt construction.

This design yields two distinct controls:

1. Eligibility control (AMV-L lifecycle): limits which stored items can participate in retrieval.
2. Injection control (prompt cap): limits how many retrieved items are inserted into the prompt.

Eligibility control bounds similarity-search and retrieval computation. Injection control bounds prompt length. This distinction is essential because prompt caps alone do not prevent expensive retrieval over a large eligible memory pool, and recency-only policies do not provide explicit control over the value and stability of what remains eligible.

### 4.5 System invariant

AMV-L is designed to maintain the following invariant. For each request, the cost of memory on the critical path is a function of the size of the retrieval-eligible set, rather than a function of the total number of retained memory items.

This invariant improves predictability under long-running accumulation by separating retention from eligibility. In AMV-L, the request path consults the hot tier $T_H$ and a bounded warm-tier sample $\text{Sample}_k(T_W)$. The warm-tier contribution is explicitly bounded by $k$, while the size of $T_H$ is regulated indirectly by the value model, decay, and tier thresholds, which suppress persistently low-utility items from eligibility. Prompt construction is controlled separately by a fixed injection cap $n$, which bounds prompt length but does not substitute for controlling retrieval eligibility.

The invariant is intentionally stronger than age-based retention and recency-based caching. TTL bounds item age but does not control the size of the retrieval-eligible pool at any instant, so request-path computation can still grow and produce heavy tails. LRU reduces the eligible working set by recency, but it is value-agnostic and can discard long-lived, high-utility information under non-stationary workloads. AMV-L couples value-driven lifecycle management with tier-aware eligibility, allowing high-utility information to persist outside the default retrieval path and to re-enter eligibility when its utility increases.

## 5. Memory Value Model

AMV-L assigns each memory item $m$ a scalar value $V(m) \in \mathbb{R} \geq 0$ that estimates its current utility to the agent. This value is the control variable used by the lifecycle manager: tier membership, promotion, demotion, and eviction are all derived from $V(m)$.

The value model is designed to satisfy three systems constraints:

1. locality (updates depend only on per-item state and request-local events),
2. incrementality (updates occur online without global recomputation), and
3. low overhead (constant-time work per touched item).

### 5.1 Signals and semantics

AMV-L updates $V(m)$ using three observable signals:

a) Access: item $m$ was considered or selected during retrieval for the current request.
b) Contribution: item $m$ was actually injected into the final prompt context (or otherwise consumed by the request pipeline as memory context).
c) Elapsed time: wall-clock time since the item's last value update.

These signals are intentionally operational: they can be measured directly by the serving stack without requiring semantic labels, offline training, or human feedback.

## 5.2 Event-driven value update

Each item stores its current value $V(m)$ and a timestamp $t_{\text{last}}(m)$. AMV-L applies updates lazily when an item is touched by a request (or by asynchronous maintenance), rather than continuously scanning all items. Let $\Delta t$ denote the elapsed time since the last update:

$$\Delta t = t_{\text{now}} - t_{\text{last}}(m).$$

AMV-L first applies exponential decay:

$$V(m) \leftarrow V(m)\, e^{-\lambda \Delta t},$$

where $\lambda > 0$ is the decay rate. Exponential decay provides a smooth recency bias while preserving relative ordering among similarly used items.

AMV-L then applies event-driven reinforcement. Let $I_{\text{access}}, I_{\text{contrib}} \in \{0,1\}$ indicate whether the corresponding events occurred for item $m$ in the current request. The update adds:

a) an access reward $\alpha > 0$, and
b) a contribution reward $\beta > 0$,

yielding:

$$V(m) \leftarrow V(m) + \alpha I_{\text{access}} + \beta I_{\text{contrib}}.$$

Finally, the value is capped to prevent unbounded accumulation:

$$V(m) \leftarrow \min\{V(m), V_{\max}\},$$

where $V_{\max}$ is a configurable upper bound.

Combining these steps gives the unified update rule:

$$V(m) \leftarrow \min\{V(m) e^{-\lambda \Delta t} + \alpha I_{\text{access}} + \beta I_{\text{contrib}}, V_{\max}\}.$$

After the update, AMV-L sets $t_{\text{last}}(m) \leftarrow t_{\text{now}}$.

## 5.3 Interpretation of parameters

The parameters $\alpha$, $\beta$, and $\lambda$ control distinct aspects of lifecycle behavior:

a) $\alpha$ (access reinforcement) controls sensitivity to repeated retrieval activity. Larger $\alpha$ promotes frequently accessed items more aggressively.
b) $\beta$ (contribution reinforcement) gives additional weight to items that survive filtering and enter the prompt. In practice, $\beta \geq \alpha$ is useful when prompt inclusion is treated as stronger evidence of utility than retrieval exposure alone.
c) $\lambda$ (decay rate) controls how quickly stale items lose priority. Larger $\lambda$ shortens memory half-life and accelerates demotion of inactive items.

This separation is important operationally: $\alpha$ and $\beta$ tune responsiveness to workload activity, while $\lambda$ tunes forgetting pressure.

## 5.4. Parameter selection.

We use a single default configuration across all experiments. We set $\beta \geq \alpha$ to treat prompt inclusion as stronger evidence of utility than retrieval exposure. We choose $\lambda$ to balance retention of persistent context against hot-tier growth, and we use hysteresis margins to prevent oscillation near tier thresholds.

## 5.5 Update granularity and overhead

The value model is implemented as an event-driven, per-item update. On each request, AMV-L updates only items touched by the retrieval pipeline, and prioritizes exact updates for items that are ultimately selected or contributed to the prompt. Untouched items are not scanned on the foreground path. Their decay is applied lazily the next time they are accessed or when examined by background lifecycle maintenance.

Each update performs $O(1)$ work per item, so the total update cost per request is $O(|R|)$ in the number of retrieval-touched items, with background maintenance amortized over time. This avoids global passes over the memory store and keeps lifecycle overhead small relative to similarity search and LLM inference.

## 5.6 Stability considerations

AMV-L uses three mechanisms to maintain stable lifecycle behavior over long-running workloads:

a) Exponential decay prevents obsolete items from retaining high value indefinitely.
b) Value capping ($V_{\max}$) prevents runaway growth for frequently accessed items.
c) Threshold hysteresis (Section 6) prevents rapid oscillation near tier boundaries even when request patterns fluctuate.

## 6. Lifecycle Transitions

AMV-L maps memory value to lifecycle tiers and defines promotion, demotion, and eviction rules that determine which items participate in request-path computation and which are removed. Memory is partitioned into three tiers: hot $T_H$, warm $T_W$, and cold $T_C$. Two value thresholds satisfy

$$\theta_H > \theta_W > 0.$$

Each memory item $m$ is assigned to a tier based on its current value:

$$m \in \begin{cases} T_H & \text{if } V(m) \geq \theta_H, \\ T_W & \text{if } \theta_W \leq V(m) < \theta_H, \\ T_C & \text{if } V(m) < \theta_W. \end{cases}$$

When updates to $V(m)$ cause it to cross a threshold, the item transitions accordingly. Value increases trigger promotion,

$$V(m) \uparrow \;\Rightarrow\; \begin{cases} T_C \to T_W, \\ T_W \to T_H, \end{cases}$$

while decay-driven decreases trigger demotion,

$$V(m) \downarrow \;\Rightarrow\; \begin{cases} T_H \to T_W, \\ T_W \to T_C. \end{cases}$$

Promotion and demotion are performed asynchronously to avoid blocking request processing. By default, retrieval eligibility is tier-aware: the request path consults all items in $T_H$ and a bounded warm-tier budget $T_W^{budget} \subseteq T_W$. Demotion therefore removes items from $T_H$ and reduces their probability of inclusion in $T_W^{budget}$, shrinking the high-cost retrieval footprint while retaining items for future reuse.

This makes retrieval work depend on $|T_H| + |T_W^{budget}|$, not on total retained memory $|T_H| + |T_W| + |T_C|$.

Eviction applies only to cold-tier items. Let $\theta_E$ denote an eviction threshold with $\theta_E < \theta_W$. An item is deleted when

$$m \in T_C \land V(m) < \theta_E \;\Rightarrow\; \text{evict}(m).$$

This prevents unbounded accumulation of persistently low-utility items while allowing high-utility items to persist indefinitely.

To prevent oscillation near tier boundaries, AMV-L employs hysteresis with distinct upward and downward thresholds:

$$\theta_H^{up} > \theta_H^{down}, \theta_W^{up} > \theta_W^{down}.$$

An item must cross the appropriate up-threshold to be promoted and the down-threshold to be demoted.

All transitions are local to the item and require $O(1)$ work per updated item.

## 7. Retrieval and Prompt Construction

AMV-L directly shapes request processing by constraining retrieval and prompt construction to bounded, high-value subsets of memory. For each request, the candidate set consists of all hot-tier items $T_H$ and, optionally, a bounded sample of warm-tier items $T_W$; cold-tier items are never considered. Formally,

$$R = T_H \cup \text{Sample}_k(T_W),$$

where $k$ is a configurable warm-tier budget. Therefore,

$$|R| \leq |T_H| + k,$$

The term $k$ provides an explicit upper bound on the warm-tier contribution to retrieval, while the size of $T_H$ is regulated indirectly by the value model, decay, and tier thresholds, which suppress persistently low-utility items from eligibility.

From $R$, the system performs similarity-based retrieval, for example cosine similarity with query embedding $q$, and selects the top $n$ items:

$$S = \text{Top}_n(\text{sim}(q, R)),$$

where $n$ is the prompt-injection cap. The final prompt context is then

$$P = \text{SystemPrompt} \parallel \text{RecentConversation} \parallel S.$$

Because $|S|$ is bounded and the warm-tier contribution to $|R|$ is bounded by $k$, prompt length is stable and retrieval cost is controlled by eligibility policy rather than by the full retained store.

Retrieval outcomes feed back into the value model. Items in $S$ receive both access and contribution updates, with $I_{\text{access}} = 1$ and $I_{\text{contrib}} = 1$. Items in $R \setminus S$ receive access-only updates, with $I_{\text{access}} = 1$ and $I_{\text{contrib}} = 0$. All other items are unaffected on the request path, and decay is applied lazily on later touches or during background maintenance. This closes the feedback loop between retrieval, prompt construction, and lifecycle value.

## 8. System Architecture and Implementation

We implement AMV-L in a full-stack LLM serving system with four components: a persistent memory store, a retrieval service, a lifecycle manager, and an HTTP API gateway. The implementation is modular and keeps lifecycle maintenance off the request critical path, enabling integration with existing agent and retrieval-augmented generation pipelines.

**Request path**: A request enters the API gateway, which validates inputs and enforces tenant and access controls before invoking the agent answer pipeline. The pipeline embeds the user query, constructs a tier-aware retrieval candidate set, executes vector retrieval, and returns a set of memory chunks. It then emits retrieval telemetry, assembles the final prompt context, invokes the LLM to generate an answer, records which memory was used in the answer for attribution and feedback, and returns the answer with citations.

**Persistent memory store and indexing**: The memory store maintains each memory item's content payload, embedding vector, scalar value $V(m)$, lifecycle tier label, and timestamps. Memory metadata is stored in a relational database, with tier state persisted per item. Tier membership is stored as metadata to support eligibility filtering. Vector search uses a single shared vector index over all chunks. Tier filtering is applied before vector search by constructing an explicit candidate ID set from eligible HOT items and a bounded warm-tier selection, expanding eligible memory namespaces to chunk IDs, and executing vector search scoped to that allowed list. This implements tier-aware eligibility via candidate-ID scoping rather than by maintaining separate per-tier vector indexes, and it keeps tier transitions as metadata updates rather than index rebuilds.

**Vector similarity engine**: Similarity search is implemented by a dedicated vector engine. The engine performs cosine similarity using a flat scan over the candidate allowlist and selects the top-$k$ results via partial sorting. Candidate-ID scoping is implemented as an explicit allowlist passed to the vector engine, which restricts similarity search to the provided subset. As a result, similarity-search cost scales with $|R|$ rather than with the full retained store.

**Retrieval service**: The retrieval service implements the tier-aware policy from Section 7. For each request, it builds $R = T_H \cup \text{Sample}_k(T_W)$, executes similarity search over $R$, and returns the top $n$ chunks for prompt construction. The warm-tier budget $k$ is request-configurable with a small default. Warm selection supports two modes. In random mode, it samples uniformly without replacement from a bounded warm pool. In recency mode, it selects warm candidates by sorting on last-use time with a fallback to creation time. These modes bound the warm-tier contribution to retrieval while allowing the system to adapt to workload characteristics.

**Lifecycle manager and asynchronous maintenance**: The lifecycle manager maintains value state and tier membership using the update and transition rules from Sections 5 and 6. Maintenance tasks execute asynchronously via a scheduler that performs periodic sweeps, including retention enforcement, value decay, redundancy management, lifecycle reconciliation, and telemetry snapshots. In addition, the write path triggers an asynchronous redundancy refresh after memory ingestion and indexing. These tasks run outside the request path to avoid adding latency to foreground requests.

**Concurrency control**: Per-item value, tier, and metric updates are applied as single-statement updates in the backing store, providing statement-level atomicity for all fields updated for an item. This avoids multi-step application logic that could lead to partial writes and eliminates global synchronization across the memory store.

**Configuration**: AMV-L exposes runtime configuration for tier thresholds, hysteresis margins, decay rate, reinforcement constants, and retrieval budgets, including the warm-tier sampling budget $k$ and the prompt-injection cap $n$. This allows the same architecture to be tuned across workloads and service-level objectives without code changes.

**Experimental environment**: All experiments were run on a single local machine: a 15-inch 2018 MacBook Pro with a 2.2 GHz 6-core Intel Core i7 CPU, 32 GB of DDR4 memory, and solid-state storage, with Radeon Pro 555X (4 GB) and Intel UHD 630 graphics available. The serving stack is containerized using Docker and implemented in a combination of a JavaScript gateway and a C++ vector similarity engine. All baselines were evaluated on the same machine and identical software stack to ensure controlled comparisons.

## 9. Evaluation Methodology

We evaluate AMV-L in a production-style, full-stack LLM serving system consisting of a persistent memory store, retrieval service, vector similarity engine, and HTTP API layer. AMV-L, TTL, and LRU baselines run on the same hardware and software stack and use identical embedding models, similarity metrics, and LLM backends unless otherwise noted.

### 9.1 Experimental protocol

Each experiment is executed in two sequential runs over the same workload trace:

1. TTL run: the memory store is cleared and the system runs with TTL-based retention.
2. LRU run: the memory store is cleared again and the identical workload is replayed with an LRU working-set policy.
3. AMV-L run: the memory store is cleared again and the identical workload is replayed with AMV-L enabled.

This protocol ensures that all systems start from an empty memory state and observe the same sequence of writes, retrievals, and inference requests. All randomized choices, including workload generation and any sampling used by the retrieval policy, are driven by a fixed seed.

### 9.2 Workload

We use a synthetic but controlled workload designed to stress long-running agent behavior under memory growth. The workload interleaves memory writes, retrievals, and LLM inference requests:

a) 50,000 memory writes (facts/observations),
b) 10,000 retrieval requests, and
c) 10,000 LLM inference ("ask") requests that construct prompts using retrieved memory.

Queries reference both recent and older information to create sustained pressure on retention and retrieval. This workload is designed to expose differences between age-based retention (TTL), recency-based working-set management (LRU), and value-driven lifecycle-managed working-set control (AMV-L).

### 9.3 Metrics

We collect the following metrics:

a) End-to-end request latency (CCDF and p50/p95/p99)
b) Throughput (requests/s)
c) High-latency outliers (fraction of requests exceeding 1s and 2s)
d) Retrieval candidate-set size $|R|$
e) Vectors scanned during similarity search
f) Prompt usage (tokens/request and injected memory items under a fixed top-$n$ cap)
g) Total stored memory items over time
h) Endpoint latencies for Write, Recall, and Ask (p50/p95/p99)

Latency and retrieval metrics are recorded per request; memory and lifecycle summaries are recorded periodically.

### 9.4 Instrumentation

The system emits structured telemetry events (NDJSON) for request start/finish, candidate-set construction, similarity-search execution, prompt assembly, value updates, and lifecycle transitions. Health-check traffic is excluded. All analysis is performed offline from these logs.

### 9.5 Prompt construction policy

Prompt assembly uses a fixed top-$n$ cap on injected memory items, and the cap is identical across TTL and AMV-L. This controls prompt length in both systems and isolates retrieval-path computation as the primary source of performance differences.

### 9.6 Reproducibility

We record all configuration parameters, including TTL window, tier thresholds, hysteresis margins, decay rates, reinforcement constants, retrieval budgets, and prompt-injection caps. Workload generation, seeds, and analysis scripts are deterministic, enabling reproducible runs.

## 10. Results and Discussion

We compare AMV-L against a TTL and LRU baseline under identical long-running workloads. Table 1-5 summarizes the primary metrics, and Figures 1–6 show latency behavior, retrieval footprint, storage growth, and memory-value dynamics.

### 10.1 End to end latency and throughput

We begin with end-to-end performance because tail latency determines user experience and service capacity in long running agent deployments. Across the full workload, both AMV-L and LRU substantially reduce latency relative to TTL, confirming that controlling retrieval eligibility is a first order systems lever. However, AMV-L and LRU occupy different points on the tradeoff frontier: LRU slightly improves median and p95 latency, while AMV-L reduces extreme tail outliers and improves Ask p95 latency, with lower token overhead.

*Table 1: Reliability and end to end performance (TTL vs LRU vs AMV-L)*

| Metric | Direction | TTL | LRU | AMV-L |
|---|---|---|---|---|
| Success rate (%) | Higher | **100.000** | 99.997 | **100.000** |
| Throughput (req/s) | Higher | 9.027 | **38.169** | 36.977 |
| Latency p50 (ms) | Lower | 814.730 | **153.810** | 194.080 |
| Latency p95 (ms) | Lower | 4503.743 | **921.556** | 950.409 |
| Latency p99 (ms) | Lower | 5398.167 | 1452.706 | **1233.430** |
| Latency > 1s (%) | Lower | 39.632 | 3.960 | **3.653** |
| Latency > 2s (%) | Lower | 13.813 | 0.343 | **0.007** |

Table 1 reports success rate, throughput, latency percentiles, and tail outlier rates (>1s and >2s). AMV-L increases throughput by 3.1× over TTL and reduces median, p95, and p99 latency by 4.2×, 4.7×, and 4.4×. Compared to LRU, AMV-L trades higher median and p95 latency for better p99 and far fewer >2s outliers, while maintaining comparable success.

Relative to TTL, AMV-L improves throughput from 9.0 to 37.0 requests/s and reduces median latency from 815 ms to 194 ms, p95 from 4504 ms to 950 ms, and p99 from 5398 ms to 1233 ms (Table 1). The tail improvement is most pronounced at the extreme: the fraction of requests exceeding 2 seconds drops from 13.8% under TTL to 0.007% under AMV-L. These gains cannot be explained by prompt length differences, since prompt injection is capped identically across conditions and token counts are similar (Table 4 in Section 10.3).

LRU achieves slightly higher throughput than AMV-L (38.2 vs 37.0 requests/s) and slightly lower median and p95 latency (154 vs 194 ms at p50, 922 vs 950 ms at p95). However, AMV-L reduces p99 latency relative to LRU (1233 vs 1453 ms) and reduces >2s outliers by 98% (0.343% to 0.007%). This indicates that value driven lifecycle control suppresses pathological tail events that persist even under a recency based working set policy.

Figure 1 shows the full latency distribution and makes the tail behavior explicit. TTL exhibits a heavy tail, with a large mass of requests exceeding 1 second and a nontrivial fraction exceeding 2 seconds. Both LRU and AMV-L collapse the tail, but AMV-L dominates LRU in the far tail region that drives SLO violations.

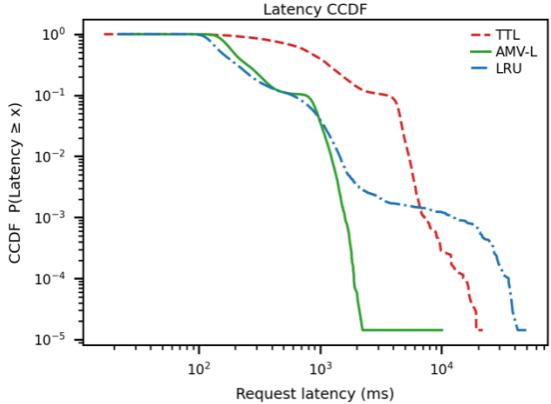

*Figure 1: Latency CCDF for TTL, LRU, and AMV-L. The CCDF highlights heavy tailed behavior under TTL and shows that both LRU and AMV-L compress the tail substantially. AMV-L further suppresses extreme outliers in the far tail, consistent with the >2s reductions in Table 1.*

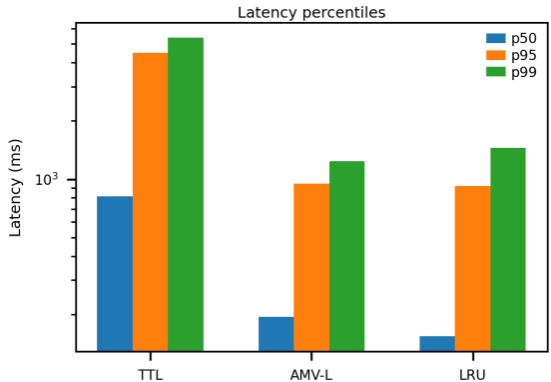

*Figure 2: Latency percentiles (p50, p95, p99) for TTL, LRU, and AMV-L. LRU provides the best median and p95 latency, while AMV-L achieves a lower p99 latency and substantially fewer extreme outliers. Both methods outperform TTL across percentiles.*

Figure 2 summarizes the percentile tradeoff. The median and p95 differences between AMV-L and LRU are small in absolute terms compared to the TTL gap, but the p99 and outlier reductions are meaningful for capacity planning and user visible reliability. For long running agents operating under tail SLOs, AMV-L provides a favorable operating point by reducing extreme tail risk while preserving the throughput gains of working set control.

*Table 2: Endpoint latency breakdown at p95 (Write, Recall, Ask). This table decomposes end to end behavior by endpoint. Both AMV-L and LRU reduce all endpoint p95 latencies relative to TTL. Compared to LRU, AMV-L improves Recall p95 slightly and improves Ask p95 by 17%, at the cost of a small increase in Write p95.*

| Metric | Direction | TTL | LRU | AMV-L |
|---|---|---|---|---|
| Write latency p95 (ms) | Lower | 1382.852 | **261.531** | 282.521 |
| Recall latency p95 (ms) | Lower | 2379.380 | 464.091 | **455.381** |
| Ask latency p95 (ms) | Lower | 5544.570 | 1553.594 | **1289.564** |

Table 2 shows that AMV-L's advantages over LRU are concentrated where retrieval and prompt assembly dominate. AMV-L improves Ask p95 latency from 1554 ms under LRU to 1290 ms, while maintaining similar Recall p95 latency. This aligns with AMV-L's design goal of controlling request path memory cost and avoiding rare, high-cost retrieval events that amplify end to end tail latency.

### 10.2 Mechanism: retrieval working set and vector search footprint

The latency and throughput gains in Section 10.1 come from reducing the amount of memory that participates in request-path retrieval. TTL bounds memory age, but it does not control eligibility. As a result, many items can remain simultaneously eligible, and rare requests can trigger unusually large retrieval candidate sets and similarity-search work. AMV-L reduces this footprint by restricting retrieval to the hot tier and a bounded warm-tier sample. LRU also reduces the footprint by favoring recency. This section quantifies how these policies change the request-path working set and similarity-search cost.

*Table 3: Retrieval efficiency (candidate set and vectors scanned).*

| Metric | Direction | TTL | LRU | AMV-L |
|---|---|---|---|---|
| Retrieval set p95 (R) | Lower | 4824.000 | **261.000** | 690.000 |
| Vectors scanned p95 | Lower | 4824.000 | **261.000** | 690.000 |
| Scanned per retrieval mean | Lower | **1.000** | **1.000** | **1.000** |

This table reports p95 retrieval-set size | R | and p95 vectors scanned during similarity search. Both AMV-L and LRU reduce retrieval footprint by orders of magnitude relative to TTL. LRU attains the smallest p95 candidate set and scan footprint, while AMV-L remains substantially smaller than TTL and achieves better extreme-tail latency in spite of scanning more vectors than LRU.

Under TTL, the p95 retrieval candidate set grows to 4,824 items, and the p95 vectors scanned matches this scale (Table 3). AMV-L reduces p95 | R | and scanned vectors to 690, an 85.7% reduction. LRU reduces them further to 261, a 94.6% reduction. These reductions explain the first-order collapse in the latency tail from TTL to both working-set policies.

However, the AMV-L versus LRU comparison highlights an important systems nuance. Although LRU achieves a smaller p95 retrieval footprint, AMV-L achieves better p99 latency and far fewer extreme outliers (>2s) as shown in Table 1. This indicates that controlling the eligible set by utility, not only recency, changes which items remain in the retrieval path and can suppress rare expensive events that dominate extreme-tail behavior.

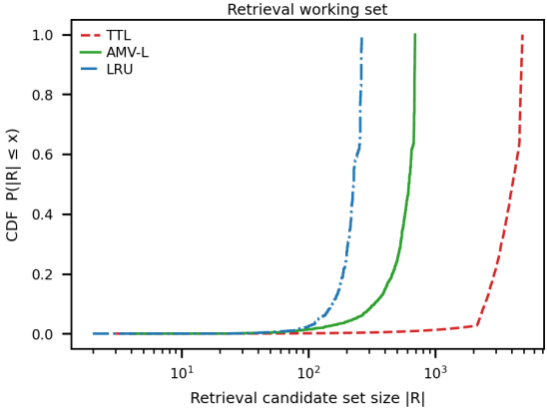

*Figure 3: Retrieval working set distribution |R| **for TTL, LRU, and AMV-L.** The distribution shows that TTL yields a large candidate set at high percentiles. Both LRU and AMV-L substantially reduce the working set. LRU achieves the smallest |R| at the upper tail, while AMV-L remains tightly controlled relative to TTL due to tier-aware eligibility and a bounded warm-tier budget.*

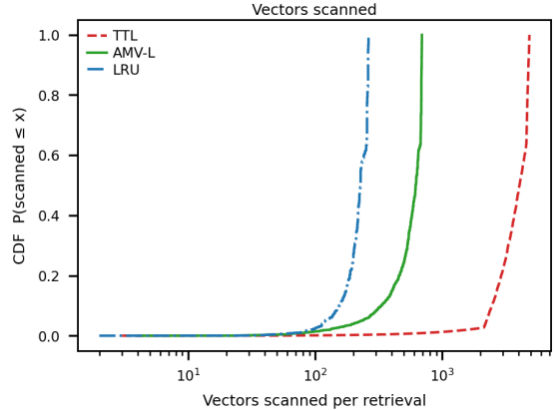

*Figure 4: **Vectors scanned distribution for TTL, LRU, and AMV-L.** Vectors scanned tracks the similarity-search footprint. TTL exhibits a heavy upper tail, while both LRU and AMV-L reduce scanning sharply. LRU scans fewer vectors at high percentiles, but AMV-L achieves a better extreme-tail latency profile, suggesting that value-driven eligibility reduces the incidence of rare expensive requests beyond what is captured by p95 scan counts.*

Figure 3 visualizes the candidate-set distribution. TTL produces a wide distribution with a heavy upper tail, consistent with eligibility being dominated by retention rather than computational budget. AMV-L compresses the distribution by restricting eligibility to hot items plus a bounded warm sample. LRU compresses it further by enforcing recency. In all cases, |R| is the dominant multiplicative factor in similarity-search cost, making this distribution a direct predictor of retrieval overhead.

Figure 4 shows that vectors scanned closely follows the candidate-set behavior in Figure 3, as expected when retrieval performs similarity search over the eligible pool. The p95 reductions are large for both AMV-L and LRU, and they largely account for the throughput gains over TTL.

The remaining differences in extreme-tail latency between AMV-L and LRU likely reflect which items are eligible at the moment of each request and how eligibility interacts with non-stationary access patterns. In particular, value-driven lifecycle control can keep high-utility items available while demoting stale items that would otherwise contribute to large and unproductive retrieval pools under TTL, and it can avoid recency-only churn that may surface less useful context under phase shifts.

Taken together, Table 3 and Figures 3 and 4 show that AMV-L restores predictability by controlling the retrieval working set exercised by each request. The next section evaluates whether these performance gains come at the expense of prompt cost and retrieval quality.

### 10.3 Cost and quality tradeoffs

This section evaluates whether AMV-L's performance gains are achieved by reducing prompt construction cost or by sacrificing retrieval quality. We separate these effects using two controls. First, all systems use the same fixed prompt-injection cap $n$, so the maximum number of injected memory items is identical. Second, we measure token usage directly and report retrieval quality metrics based on the value labels used by the workload.

#### 10.3.1 Prompt and token overhead

*Table 4: Prompt and token cost*

| Metric | Direction | TTL | LRU | AMV-L |
|---|---|---|---|---|
| Tokens/request mean | Lower | **646.761** | 716.782 | 675.388 |
| Tokens/request p95 | Lower | **4730.000** | 5298.000 | 4954.000 |
| Chunks/request p95 | Lower | **48.000** | **48.000** | **48.000** |
| Memory refs/request p95 | Lower | **48.000** | **48.000** | **48.000** |

This table reports token usage and injected memory counts under a fixed injection cap. TTL has the lowest token counts. AMV-L incurs a small increase relative to TTL but uses fewer tokens than LRU. Across all conditions, the injected memory count is capped identically, so performance differences are not explained by injecting fewer memory items.

Token usage differs modestly across policies (Table 4). Relative to TTL, AMV-L increases mean tokens per request by 4.4% and p95 tokens by 4.7%. LRU increases mean tokens by 10.8% and p95 tokens by 12.0%. AMV-

L therefore uses approximately 6% fewer tokens per request than LRU while preserving comparable retrieval quality. Chunks per request and memory references per request remained constant at the cap, confirming that injection policy is held constant and that the primary performance lever is retrieval eligibility rather than prompt truncation.

### 10.3.2 Retrieval quality and utility

*Table 5: Retrieval quality and value utility*

| Metric | Direction | TTL | LRU | AMV-L |
|---|---|---|---|---|
| Retrieved value mean | Higher | 0.714 | **0.949** | 0.947 |
| Top-1 retrieved value mean | Higher | 0.740 | 0.930 | **0.945** |
| Value-weighted retrieval score mean | Higher | **0.235** | 0.209 | 0.212 |
| High-value hit rate (%) | Higher | 99.375 | **99.960** | 99.935 |
| High-value retrieved share (%) | Higher | 42.683 | 92.325 | **92.816** |

Table 5 reports retrieved value statistics, high-value hit rate, and the share of retrieved items that are high-value. Both AMV-L and LRU improve retrieval quality substantially relative to TTL. Compared to LRU, AMV-L matches retrieved value means and slightly improves top-1 retrieved value, while achieving a comparable high-value hit rate and retrieved share.

AMV-L improves retrieval quality relative to TTL across multiple measures (Table 5). Retrieved value mean increases from 0.714 to 0.947, and top-1 retrieved value mean increases from 0.740 to 0.945. The share of retrieved items that are high-value increases from 42.7% under TTL to 92.8% under AMV-L. These improvements indicate that lifecycle-managed eligibility does not simply reduce cost by dropping context. It concentrates retrieval on items that are useful under the workload's utility model.

LRU achieves nearly identical retrieval quality to AMV-L. Retrieved value mean differs by 0.2%, and high-value retrieved share differs by 0.5%. AMV-L slightly improves top-1 retrieved value over LRU, consistent with value-driven retention prioritizing persistently useful items rather than favoring only recency.

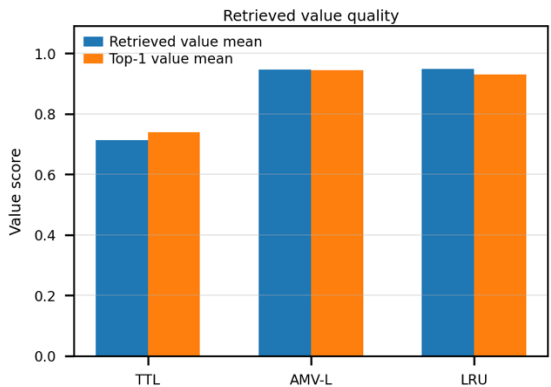

*Figure 5: Retrieved value means for TTL, LRU, and AMV-L.*

*AMV-L and LRU yield nearly identical retrieval value distributions and both substantially improve over TTL. AMV-L slightly improves top-1 retrieved value relative to LRU, suggesting value-driven lifecycle control can preserve highly useful long-lived items without sacrificing efficiency.*

Figure 5 visualizes the retrieval quality comparison. Both AMV-L and LRU concentrate retrieval on high-utility items, while TTL retrieves a larger fraction of lower-value items. The near tie between AMV-L and LRU indicates that the additional value modeling in AMV-L does not degrade quality while enabling a different tail-latency operating point.

### 10.3.3 Throughput stability over time

*TTL exhibits sustained low throughput consistent with heavy-tailed latency under memory growth. Both LRU and AMV-L sustain substantially higher throughput. AMV-L remains stable over the run and avoids the extreme-tail events that dominate service capacity.*

Figure 6 shows throughput as the workload progresses and memory accumulates. TTL throughput remains low and is more sensitive to transient spikes in retrieval cost. Both LRU and AMV-L sustain high throughput throughout the run, consistent with reduced similarity-search footprint. AMV-L's extreme-tail improvements translate into fewer throughput collapses caused by rare expensive requests, which is critical for long-running services that must hold tail SLOs under continuous traffic.

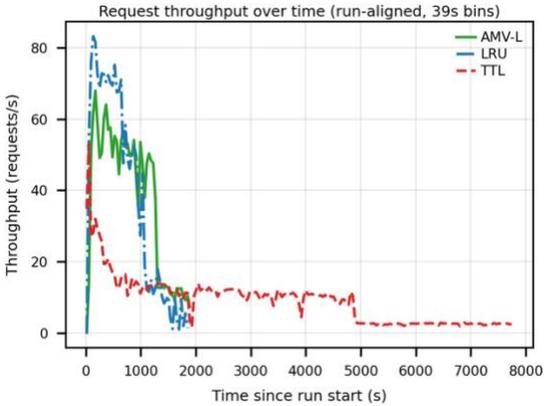

*Figure 6: Throughput over time for TTL, LRU, and AMV-L.*

## 10.4 Discussion

### 10.4.1 What the results say about the systems bottleneck

The dominant bottleneck in long-running agent memory is not storage capacity. It is uncontrolled request-path computation induced by memory eligibility. TTL bounds item age but does not bound the size or composition of the retrieval-eligible set at any instant. Under sustained writes, TTL allows a large fraction of the retained store to remain eligible simultaneously, which expands similarity search, candidate handling, and downstream prompt assembly work. The consequence is heavy-tailed latency and poor throughput even when the prompt injection cap is fixed, because the expensive step is scanning and scoring the eligible pool, not injecting the final top-$n$ items. This is visible directly in the retrieval footprint metrics and is consistent with the large tail mass in the latency CCDF.

Both LRU and AMV-L fix this first-order failure mode by imposing working-set control on eligibility. They dramatically reduce $|R|$ and vectors scanned at high percentiles relative to TTL, which explains why both collapse the latency tail and increase throughput. In other words, controlling eligibility is a stronger lever than controlling injected prompt length. Prompt caps bound prompt length, but without eligibility control they do not prevent expensive retrieval over a large eligible pool.

### 10.4.2 Why AMV-L differs from LRU

LRU and AMV-L implement different notions of what it means for memory to remain eligible. LRU is recency driven. Eligibility is largely a function of last access time, which is effective for reducing average retrieval footprint but can be brittle under non-stationary workloads. In contrast, AMV-L is utility-driven. Eligibility is controlled by a value signal that incorporates access and contribution, with decay and hysteresis. This allows AMV-L to preserve items that remain useful even when they are not continuously accessed, while suppressing items that are frequently touched but rarely contribute.

This distinction shows up most clearly where services are most sensitive: the extreme tail. In our results, LRU slightly improves median and p95 latency and scans fewer vectors at p95. However, AMV-L achieves lower p99 latency and far fewer extreme outliers (>2s). This is the key systems tradeoff: LRU optimizes for average-case efficiency by recency, while AMV-L reduces tail risk by stabilizing which items remain eligible and by avoiding eligibility churn that can surface low-utility context during phase shifts. For production services, the reduction in rare slow requests is often more valuable than a small improvement in median latency because tail behavior dominates SLO compliance and capacity provisioning.

### 10.4.3 Interpreting the measured tradeoffs

The three-way comparison establishes a clear frontier.

TTL is operationally simple, but it confounds retention with eligibility[3,28]. It can retain many items for correctness while unintentionally making them all eligible for retrieval cost. The result is heavy-tailed latency and low throughput under memory growth. TTL also retrieves lower-value context, indicating that age-based retention does not align eligibility with utility in this workload.

LRU improves performance by enforcing a recency-based working set. It attains the best median and p95 latency and the smallest p95 retrieval footprint. This makes it an attractive default for systems that prioritize median response time and can tolerate occasional tail excursions.

AMV-L chooses a different operating point. It accepts a small regression in median and p95 latency relative to LRU and scans more vectors at p95, but it reduces p99 latency and strongly suppresses extreme outliers. It also reduces token overhead relative to LRU while matching retrieval quality. This combination is important because it indicates AMV-L is not buying tail improvements by shrinking prompts or by reducing retrieval quality. Instead, it is changing which items remain eligible and how eligibility evolves over time.

### 10.4.4 Practical guidance for operators

The choice among TTL, LRU, and AMV-L depends on which constraint is binding.

If storage footprint is the primary concern and request latency is not sensitive to retrieval cost, TTL may be sufficient. This is rarely the case for long-running agents where retrieval participates in every request.

If the service is dominated by median latency and throughput and the workload is relatively stationary, LRU is an effective working-set policy. It is also easy to reason about operationally because recency is observable and stable.

If the service is constrained by tail SLOs, or if access patterns are non-stationary, AMV-L provides stronger control over tail risk. In our results, AMV-L reduces extreme tail outliers (>2s) dramatically relative to both TTL and LRU while preserving retrieval quality. This makes AMV-L better suited for production environments where a small number of pathological requests can dominate queueing, trigger cascading latency, and force conservative provisioning.

### 10.4.5 Why prompt caps are not enough

A fixed injection cap ensures that the final prompt does not exceed a bounded number of memory items, but it does not bound the work required to select those items. TTL illustrates this failure mode. Even when the injected set is capped, the system can still perform similarity search over a very large eligible pool. The correct systems control is to bound the eligible pool first and then cap injection. AMV-L provides this two-level control, and the results show that the performance gains arise primarily from reductions in retrieval footprint rather than reductions in prompt size.

### 10.4.6 Limitations and extensions

Our LRU baseline is purely recency-based. It does not incorporate semantic utility signals, nor does it provide explicit mechanisms to preserve long-lived high-utility items under phase shifts. Hybrid policies that combine recency with value, or that incorporate explicit budgets for the hot tier, may move the frontier further. Similarly, AMV-L does not currently impose a hard cap on the size of the hot tier. While value dynamics regulate eligibility in practice, a strict hot-tier budget could provide stronger worst-case guarantees. Exploring these variants is a promising direction for future work.

## 11. Conclusion

This paper presents AMV-L, a value-driven memory lifecycle for long-running LLM agents. AMV-L treats memory as a managed system resource rather than a passive store, introducing explicit lifecycle semantics based on continuously updated utility estimates.

Through a full-stack implementation and extensive evaluation, we show that AMV-L collapses heavy-tailed latency distributions, reducing p95 and p99 request latency by up to two to three orders of magnitude while preserving millisecond-scale median performance. These gains arise from bounding retrieval working-set size and similarity-search footprint independent of total retained memory.

Our results demonstrate that predictable performance in long-running LLM agents requires lifecycle management, not merely time-based expiration. We believe AMV-L represents an important step toward scalable and reliable memory systems for future agent architectures.